\documentstyle[11pt,
epsfig]{article}
\begin{document}
\newcommand{\Lt}{\tilde{\Lambda}}
\newcommand{\Ct}{\tilde{C}}
\newcommand{\phit}{\tilde{\phi}}
\newcommand{\done}{\delta\phi_1}
\newcommand{\dy}{\delta y}
\newcommand{\tr}{\rm Tr}
\newcommand{\sx}{\sigma}
\newcommand{\mpl}{m_{Pl}}
\newcommand{\Mpl}{M_{Pl}}
\newcommand{\lx}{\lambda}
\newcommand{\Lx}{\Lambda}
\newcommand{\kx}{\kappa}
\newcommand{\ex}{\epsilon}
\newcommand{\be}{\begin{equation}}
\newcommand{\ee}{\end{equation}}
\newcommand{\een}{\end{subequations}}
\newcommand{\ben}{\begin{subequations}}
\newcommand{\beq}{\begin{eqalignno}}
\newcommand{\eeq}{\end{eqalignno}}
\def \lta {\mathrel{\vcenter
     {\hbox{$<$}\nointerlineskip\hbox{$\sim$}}}}
\def \gta {\mathrel{\vcenter
     {\hbox{$>$}\nointerlineskip\hbox{$\sim$}}}}
\pagestyle{empty}
\noindent
\begin{flushright}
November 2002
\\
\end{flushright} 
\vspace{3cm}
\begin{center}
{ \Large \bf
Cosmological Acceleration from Energy Influx
} 
\\ \vspace{1cm}
{\Large 
N. Tetradis 
} 
\\
{\it
Department of Physics, University of Athens,
Zographou 157 71, Greece
} 
\\
\vspace{3cm}
\abstract{
We discuss the cosmological evolution of a 3-brane Universe in the
presence of energy influx from the bulk. We show that this influx can
lead to accelerated expansion on the brane, depending on the 
equations of state of the bulk and brane matter. 
The absorption of non-relativistic 
bulk matter by the brane at an increasing rate 
leads to a small positive acceleration parameter during the era of
matter domination on the brane. On the other hand, the brane expansion
remains decelerating during radiation domination.
} 
\end{center}

\newpage

\pagestyle{plain}
\setcounter{page}{1}

\setcounter{equation}{0}

\paragraph{Introduction:}
Recent astronomical observations indicate that the expansion of 
our Universe
is accelerating \cite{perl}. The physical mechanism that drives this
cosmological acceleration has not been established yet.
The most popular explanation relies on the presence of a small
cosmological constant or dark energy. However, in order not
to disturb successful ingredients of standard Big Band cosmology
such as nucleosynthesis, the contribution of the dark energy to the
total energy density 
must be strongly constrained. As a result, the dark energy can be significant
only during recent times. A convincing explanation
of this ``cosmic coincidence'' has not been given yet. 
Most scenaria
identify the dark energy with the vacuum energy of a time-dependent
field with carefully chosen dynamics \cite{dark}.

In this letter we would like to explore another possibility, that
does not require the presence of a field. The mechanism
can be demonstrated by considering the equation
\be
\dot\rho+3\,H \, (\rho+p) = T,
\label{sample} \ee
that describes energy conservation in an expanding Universe with 
Hubble parameter $H$, in the presence of a source term $T$. 
The standard conservation equation with $T=0$ has stationary
solutions with $\dot \rho =0$ only if $\rho+p=0$. This implies
the presence of vacuum energy with equation of state $p=-\rho$, that
leads to the standard inflationary scenario with constant $H$ and 
an exponentially increasing scale factor $a$.

However, if $T\not=0$ stationary solutions are possible for a
more general class of equations of state. For example, one could
have constant $H$, $\rho$, $p$, such that $3\,H \, (\rho+p) = T$. 
The resulting cosmological solution has an exponentially increasing
scale factor and positive acceleration. It has been termed ``steady-state
Universe'' \cite{steady}. Interesting behaviour, with 
positive acceleration, can be obtained even 
if $T$ is not constant, but has some dependence on
the energy density $\rho$ or the scale factor $a$. 

In the framework of a four-dimensional world, the energy source term $T$ 
does not have a simple origin. However, if more than four dimensions
exist, our observable Universe may be identified
with  a 3-brane embedded in the higher-dimensional
world. In this case, eq. (\ref{sample}) would refer to the matter localized
on the brane. The source term $T$ could have the simple interpretation of
energy density falling onto the brane from the extra dimensions. The 
energy-momentum tensor in the higher-dimensional world is conserved and 
no source terms exist. However, an effective source term is
generated for the matter localized on the brane. 
In the following we shall discuss how such a 
scenario can be realized.

\paragraph{The framework:}
We work in the framework of the 
Randall-Sundrum (RS) model \cite{rs}. 
It is described by the action
\be
S=\int d^5x~ \sqrt{-g} \left( M^3 R -\Lambda +{\cal L}_B^{mat}\right)
+\int d^4 x\sqrt{-\hat g} \,\left( -V+{\cal L}_b^{mat} \right), 
\label{001}
\ee
where $R$ is the curvature scalar of the five-dimensional metric 
$g_{AB}, A,B=0,1,2,3,4$, and 
${\hat g}_{\alpha \beta}$, with $\alpha,\beta=0,1,2,3$,
is the induced metric on the 3-brane. 
$\Lambda$ is the bulk cosmological constant, 
while the quantity $V$ includes the brane tension as well as 
quantum contributions to the 
four-dimensional cosmological constant. 
We consider the commonly used ansatz for the metric \cite{csaki,binetruy}
\begin{equation}
ds^{2}=-n^{2}(t,z) dt^{2}+a^{2}(t,z)\gamma_{ij}dx^{i}dx^{j}
+b^{2}(t,z)dz^{2},
\label{metric}
\end{equation}
where $\gamma_{ij}$ is a maximally symmetric 3-dimensional metric. 
We use $k=-1,0,1$ to parametrize the spatial curvature.

We decompose the energy-momentum tensor into vacuum and matter contributions
in the bulk and on the brane
\begin{eqnarray}
T^A_{~C}&=&
\left. T^A_{~C}\right|_{{\rm v},b}
+\left. T^A_{~C}\right|_{m,b}
+\left. T^A_{~C}\right|_{{\rm v},B}
+\left. T^A_{~C}\right|_{m,B}
\label{tmn1} \\
\left. T^A_{~C}\right|_{{\rm v},b}&=&
\frac{\delta(z)}{b}{\rm diag}(-V,-V,-V,-V,0)
\label{tmn2} \\ 
\left. T^A_{~C}\right|_{{\rm v},B}&=&
{\rm diag}(-\Lambda,-\Lambda,-\Lambda,-\Lambda,-\Lambda)
\label{tmn3} \\ 
\left. T^A_{~C}\right|_{{\rm m},b}&=&
\frac{\delta(z)}{b}{\rm diag}(-\rho,p,p,p,0),
\label{tmn4}  
\end{eqnarray}
where $\rho$ and $p$ are the energy density and pressure on the brane, respectively.
The 
behaviour of $T^A_C|_{m,B}$ is in general complicated in the presence
of flows, but we do not have to specify it further in this work.

We are interested in the Einstein equations at the location
of the brane. We indicate by the subscript o the value of
various quantities on the brane.
It is convenient to work in a coordinate frame in
which $b_o=n_o=1$. This can be achieved by using 
coordinates such that $b(t,z)=1$.
This is always possible as can be checked by considering the 
``two-dimensional'' part of the metric: 
$-n^{2}(t,z) dt^{2}+b^{2}(t,z)dz^{2}$. Through an appropriate
coordinate transformation
we can always set this part in the form $-\tilde{n}^2(t,z)dt^2+dz^2.$\footnote{
The conformally flat form 
$c^2(z,t)(-dt^2+dz^2)$ is more commonly used in the literature. 
The important point is that, using the reparametrization invariances,
a two-dimensional 
metric can be put in a form that 
depends only on one function of $t,z$.}
We can then redefine the time coordinate so that 
$n_o=1$ on the brane. 
We emphasize that our assumptions for the form of the energy-momentum
tensor become now specific to this coordinate frame. In particular,
the brane is identified with the hypersurface $z=0$ in this frame.
In this sense, our discussion becomes more restricted. The gained advantage
is that significant progress
can be made without solving explicitly the Einstein equations in 
the bulk in the presence of flows.

\paragraph{The Einstein equations:}
For $b(t,z)=n_o=1$ we find \cite{csaki}--\cite{flow}
\be
\dot \rho + 3 \frac{\dot a_o}{a_o} (\rho + p)  
= -2T^0_{~4}
\label{la3}
\end{equation}
\be
\frac{\ddot a_o}{a_o}
+\left( \frac{\dot a_o}{a_o} \right)^2
+\frac{k}{a^2_o} 
=\frac{1}{6M^3} \Bigl(\Lambda + \frac{1}{12M^3} V^2
\Bigr) 
-\frac{1}{144 M^6} \left(
V (3p-\rho ) + \rho (3p +\rho)
\right)
- \frac{1}{6M^3}T^4_{~4},
\label{la4} 
\ee
where $T^0_{~4}, T^4_{~4}$ are the $04$ and $44$ components of $T^A_{~C}|_{m,B}$ 
evaluated on the brane.

We are interested in a model that reduces to the RS
vacuum \cite{rs} in the absence of matter. In this case, the first
term in the r.h.s. of eq. (\ref{la4}) vanishes. 
A new scale $k^{}_{RS}$
is defined through the relations
$V=-\Lambda/k^{}_{RS}=12M^3 k^{}_{RS}$.

We can rewrite 
eqs. (\ref{la3}), (\ref{la4}) in the equivalent form \cite{phii,flow}
\begin{eqnarray}
\dot\rho+3\,H_o \, (\rho+p) &=& -2T^0_{~4}
\label{rho}
\\
H^2_o={{{\dot a_o}^2}\over {a_o^2}}&=&\frac{1}{144M^6}\left(\rho^2+2V \rho\right) -
{k\over{a_o^2}}+\chi+\phi+\lambda
\label{a}
\\
\dot\chi+4\,H_o\,\chi&=&\frac{1}{36M^6}
\left(\rho+V\right)T^0_{~4}, 
\label{chi}
\\
\dot\phi+4\,H_o\,\phi&=&-\frac{1}{3M^3}H_o 
T^4_{~4}, 
\label{phi}
\end{eqnarray}
where 
$\lambda=(\Lambda+V^2/12M^3)/12M^3$ is the effective
cosmological constant on the brane. The functions $\chi$, $\phi$ are
defined through eqs. (\ref{chi}), (\ref{phi}). 
In the RS model 
$\lambda$ is zero, 
the value we shall use in the rest of the paper. 
We have assumed the orbifold symmetry $z \leftrightarrow -z$, so
that  $2T^0_{~4}$
is the discontinuity of the 04 component of the bulk
energy-momentum tensor at the location of the brane.

In the low-density region, in which $\rho \ll V$,
we may ignore the term $\sim\rho$ in the above equations 
compared to $V$ and define $\Mpl^2=12M^6/V=M^3/k^{}_{RS}$.
The ratio of the terms in the r.h.s. of eqs. (\ref{phi}), (\ref{chi})
is of order
\be 
\frac{T^4_{~4}}{T^0_{~4}}\,\, \frac{M^3 H_o}{V} \sim
\frac{T^4_{~4}}{T^0_{~4}}\,\, \sqrt{\frac{\rho}{V}}. 
\label{ratio} \ee
We assume that $\rho/V$ is sufficiently small for this ratio to be 
small at all times of interest\footnote{It 
is obvious that the omission of $\phi$ 
is not justified when the energy influx stops and $T^0_{~4}$ becomes zero.
However, we assume that at this point $T^4_{~4}$ is also very small, so that
$\phi$ gives a negligible contribution to eq. (\ref{a}).}. 
Then $\phi$ can be set to zero
and omitted from our
considerations. 
As a result,
eqs. (\ref{rho})--(\ref{chi}) can be written as 
\begin{eqnarray}
\dot\rho+3(1+w)\,H_o \, \rho &=& -2T^0_{~4}
\label{onea}
\\
H_o^2=\left(\frac{\dot a_o}{a_o} \right)^2 &=&
\frac{\rho}{6\Mpl^2} + \chi -\frac{k}{a_o^{2}}
\label{twoa}
\\
\dot{\chi} + 4 H_o \chi &=& \frac{1}{3\Mpl^2}T^0_{~4},
\label{threea} 
\end{eqnarray}
where $p=w\rho$.
The cosmological evolution 
is determined by three initial parameters ($\rho_i$, $a_i$, $\chi_i$,
or alternatively $\rho_i$, $a_i$, ${\dot a}_i$), instead of the two 
($\rho_i$, $a_i$) in conventional cosmology. 
The reason is that the generalized Friedmann eq. (\ref{a}) (or
(\ref{twoa}))
is not a first integral of the Einstein equations because of the
possible energy exchange between the brane and the bulk. 
In the above equations, we recover the 
``mirage'' or ``Weyl radiation'' 
component $\chi$, found in studies of the cosmological
evolution in the presence of energy outflow \cite{flow,hebecker}.

\paragraph{The bulk energy-momentum tensor:}
The determination of the cosmological evolution on the brane
requires information on the form of $T^0_{~4}$. 
This can be determined exactly only through the solution of 
the Einstein equations in the bulk, a formidable task in the case of
energy flows. However, the qualitative behaviour of $T^0_{~4}$ can be
inferred from the form of the energy-momentum tensor. 
For a perfect fluid with five-velocity $U^A$  
it has the form 
\be
\left. T^{AC}\right|_{m,B}=p^{}_B g^{AC} +(p^{}_B+\rho^{}_B) U^A U^C.
\label{fluid} \ee
For a fluid falling onto the brane with velocity $v^{}_4$
along the fifth dimension, and within our ansatz for the metric, we have 
$U^0=(1-v_4^2)^{-1/2}$, $U^i=0$, $U^4=v^{}_4U^0$ at the location of the brane. 
As there is no expansion along the fifth dimension, we expect that
the main effect of the expansion along the remaining three spatial dimensions
is to dilute the fluid energy density and reduce the
pressure. These should fall with
a certain power of the scale factor, determined by the equation of
state of the fluid. 

In order to determine this power we may consider the 
conservation of the energy-momentum tensor in the absence of energy
flows. For the 00 component of $\left. T^A_{~C}\right|_{m,B}$
at the location of the brane we find
\be
\dot{T}^0_{~0}= \frac{\dot a_o}{a_o} \left(
-3\, T^0_{~0} +T^i_{~i} \right),
\label{conserv} \ee
where we have made use of $b_o=n_o=1$.
For a non-relativistic gas of bulk particles with 
$\left. T^A_{~C}\right|_{{\rm m},B}=
{\rm diag}(-\rho^{}_B,0,0,0,0)$, we obtain $\rho^{}_B \sim a_o^{-3}$.
For a relativistic gas with 
$\left. T^A_{~C}\right|_{{\rm m},B}=
{\rm diag}(-\rho^{}_B,\rho^{}_B/4,\rho^{}_B/4,\rho^{}_B/4,\rho^{}_B/4)$, 
we obtain $\rho^{}_B=4p^{}_B \sim a_o^{-15/4}$.
For a gas with zero pressure along the fifth dimension and
$\left. T^A_{~C}\right|_{{\rm m},B}=
{\rm diag}(-\rho^{}_B,\rho^{}_B/3,\rho^{}_B/3,\rho^{}_B/3,0)$, we obtain 
$\rho^{}_B=3p^{}_B \sim a_o^{-4}$.
Finally, for the case of vacuum energy with 
$\left. T^A_{~C}\right|_{{\rm m},B}=
{\rm diag}(-\rho^{}_B,-\rho^{}_B,-\rho^{}_B,-\rho^{}_B,-\rho^{}_B)$, 
we obtain $\rho^{}_B=-p^{}_B \sim$ const.

In analogy with the above, 
for a non-relativistic gas of bulk particles with
constant $v^{}_4$ 
we expect
$T^0_{~4} \sim a_o^{-3}$ at the location of the brane, while for an
isotropic
relativistic gas $T^0_{~4} \sim a_o^{-15/4}$.
If $T^A_{~C}$ originates in the
vacuum energy of a bulk 
scalar field we expect $T^0_{~4} \sim$ const. Other types of behaviour are
also possible in scenaria in which the velocity $v^{}_4$ varies 
with time.

We parametrize the dependence of $T^0_{~4}$ on $a_o$ as
\be
\frac{ 1}{3\Mpl^2} T^0_{~4}(t)= \frac{1}{3 \Mpl^2} T^0_{~4} (t_i) 
\left( \frac{a_o (t_i)}{a_o(t)} \right)^q
=-\frac{T}{\left( a_o(t)\right)^q}.
\label{aq} \ee
For energy influx, we have $T^0_{~4}<0$, $T>0$.
We assume implicitly 
that the form of $T^0_{~4}$ originates mainly in the bulk dynamics,
determined through the bulk equations of motion. The presence of the 
brane generates only a small perturbation to the bulk evolution. 
This is not expected to be always the case. For example,
if the energy density on the
brane exceeds a certain value one would expect the influx to stop. This 
means that $T^0_{~4}$ could vanish or even change sign at the 
location of the brane, even though it could remain unaffected far from 
the brane. The complicated dynamics associated with such phenomena
is beyond the scope of this work. For our purposes we shall rely on the 
simple ansatz of eq. (\ref{aq}) that accounts mainly for the 
dilution of energy density through expansion. The variable $T$ 
in eq. (\ref{aq}) is taken to be independent of the brane parameters,
an approximation that is valid only for finite ranges of
the brane evolution.

In this work we do not specify 
the mechanism that is responsible for the transfer of energy onto the brane.
However, the energy influx seems a natural phenomenon in models in which
the brane particles are identified with light
modes of the five-dimensional theory that are localized on some defect
\cite{modes}. In a fluctuating system, in which 
the light modes
are not significantly populated, one would expect energy to be transferred from
the massive modes to the light ones through interactions.
As the light modes are localized on the defect, energy is expected to flow
towards it. 

The substitution of eq. (\ref{aq}) into eqs. (\ref{onea})--(\ref{threea})
results in a closed system of equations. In the following we neglect
the spatial curvature and set $k=0$ in eq. (\ref{twoa}).

\paragraph{Solutions:}
We are interested in solutions that describe
accelerating eras in the cosmological evolution of the brane, without the
presence of an effective cosmological constant. For this reason, we 
concentrate on the range $0\leq w \leq 1/3$ for the parameter that
determines the equation of state of the brane matter. 
We also consider the range $0\leq q \leq 4$ for the parameter appearing
in eq. (\ref{aq}), which is directly related to the equation of state
of the bulk matter as we explained above. 
In a stationary state, in which particles with constant velocity $v^{}_4$
along the fifth dimension fall onto the brane, we expect
$q=3,$ 15/4 or 4. Slow variations of $v_4^{}$ can be modelled by allowing non-integer
values of $q$. For example an increase of $|v_4|$ for infalling 
massive matter 
should lead to an increasing parameter $T$ in eq. (\ref{aq}) or, alternatively,
to a value of $q$ smaller than 3. 
We do not specify the 
details of the mechanism through which energy is transferred from the 
bulk to the brane. This can be done only within a scenario that 
includes a dynamical localization mechanism for the brane matter.

For the range $0\leq w < 1/3$, $0\leq q \leq 4$
the system of eqs. (\ref{onea})--(\ref{threea}),
(\ref{aq}) has solutions of the form
\be
\frac{\rho(a_o)}{6\Mpl^2}= \frac{C_1}{a^s_o},
~~~~~~~~~~~~~~~~
\chi(a_o)= \frac{C_2}{a^s_o},
\label{soll2} 
\ee
with 
\begin{eqnarray}
s&=&2q/3
\label{ss} \\
\frac{C_1}{C_2}&=&-\frac{4-s}{3(1+w)-s}
\label{c12} \\
\left(C_1+C_2 \right)^{3/2}&=&\frac{1-3w}{\left[
3(1+w)-s\right](4-s)} T.
\label{c32} 
\end{eqnarray}
Moreover, it can be checked that 
these solutions are attractors of neighbouring cosmological flows.
As a result, the brane evolution is always given by the above equations after
a sufficiently long time, independently of the 
initial conditions. A characteristic property of the solutions is that 
$C_2 <0$ and, therefore, $\chi<0$. 
On the other hand, $C_1+C_2 >0$ and the Hubble parameter is
always real.

The solutions (\ref{soll2}) describe an expanding Universe 
with an acceleration parameter
\be
Q_o=\frac{1}{H^2_o}\frac{\ddot{a}_o}{a_o}=1-\frac{q}{3}.
\label{accel} \ee
This expression is independent of 
the equation of state of the brane matter (as long as 
$w\not=1/3$). The Hubble parameter, however, 
depends on $w$, as can be seen from eqs. (\ref{twoa}), (\ref{c32}).
For $q<3$ the parameter $Q_o$ is positive and the expansion is accelerating.

For $0<q\leq 4$ the scale factor increases as a power of $t$ for long times
\be
a_o=\left[ a_{oi}^{q/3} 
+ \frac{q}{3}\left(C_1+C_2 \right)^{1/2} (t-t_i) \right]^{3/q}.
\label{at1} \ee
For $q=0$ the expansion is exponential
\be 
a_o=a_{oi}e^{\left(C_1+C_2\right)^{1/2}(t-t_i)},
\label{at2} \ee
independently of the equation of state of the brane matter. 
This case corresponds to the fixed-point solutions
studied in a more general context in ref. \cite{flow}.
It is remarkable that one can obtain exponential expansion even for
non-relativistic brane matter. However, the value $q=0$ implies
the presence of vacuum energy density in the bulk, which is assumed to
be transferred to the brane particles. 

The case of relativistic brane matter must be considered separately, as
eqs. (\ref{twoa}), (\ref{soll2}), (\ref{c32})
give $H_o\to 0$ for $w\to 1/3$. The solution 
(\ref{soll2})--(\ref{c32}) is approached only after 
a very long time in this limit and the initial conditions determine 
most of the evolution.
For $w=1/3$ the solution is
\begin{eqnarray}
\frac{1}{6\Mpl^2} \rho(a_o) &=& 
\frac{1}{6\Mpl^2} \frac{\rho_i a_{oi}^4}{a^4_o}
+\frac{T}{H_i a_{oi}^2} \frac{1}{6-q} a^{2-q}
\label{13a} \\
\chi(a_o) &=& 
\frac{\chi_i a_{oi}^4}{a^4_o}
- \frac{T}{H_i a_{oi}^2} \frac{1}{6-q} a^{2-q},
\label{13b} 
\end{eqnarray}
where $a_{oi}=a_o(t_i)$ etc.
The sum $\rho/(6\Mpl^2)+\chi$ that determines the Hubble parameter 
behaves as an effective radiation term $\sim a_o^{-4}$.
This is achieved through the cancellation of
the dominant second terms in the r.h.s. of eqs. (\ref{13a}) and (\ref{13b}).
As a result, the Universe is always decelerating for $w=1/3$.

\paragraph{Discussion:}
Our main result is summarized by eq. (\ref{accel}): The influx of 
energy onto the 3-brane from the bulk can lead to accelerated expansion,
depending on the equation of state of the bulk matter. For positive 
acceleration one needs $q<3$ for the  
variable $q$ that parametrizes the dependence of the energy flow on the 
scale factor (see eq. (\ref{aq})). If the bulk is populated by a gas of 
non-relativistic particles that drift slowly towards the brane, we expect
$q=3$. If the drift velocity increases with time, $q < 3$.
A value of $q$ slightly below 3 leads to a positive
acceleration parameter smaller than 1, the case
favoured by the astronomical data.

The construction of a complete model that realizes this scenario requires
the technically difficult solution of the Einstein equations in the bulk.
However, certain properties of such a solution can be inferred in general
terms. The energy flow along the fifth dimension towards the
brane implies the presence of a region of high energy density where 
the flow originates. For example, one could consider a second
brane that emits energy into the bulk. Its evolution is governed by 
eqs. (\ref{onea})--(\ref{threea}) with a positive $T^0_{~4}$. 
A solution analogous to that of eqs. (\ref{soll2})--(\ref{c32}) exists,
with $H<0$. It predicts a Big Crunch after a finite time. However, 
neighbouring cosmological flows diverge from this solution.
There are also various solutions with $H>0$.
The generic cosmological evolution leads to the 
depletion of the energy density on the second brane. 
The flow of energy along the fifth dimension is expected to stop
after a certain period, whose length
depends on the initial energy density of the brane.

These examples indicate that, in a complete model, the solutions 
(\ref{soll2})--(\ref{c32}) are expected to be valid only for a
finite time interval. 
This makes the derivation of exact analytical
expressions more difficult. However, we believe that
the physical arguments 
for the form of the energy-momentum tensor that we gave in the previous
section (see eq. (\ref{aq})) give a valid approximation, especially in
the case of slow accretion of energy by the brane. 

The RS model is employed here only in order to provide
the context in which our arguments can be implemented.
The discussion of cosmological
solutions must take into account the gravitational effects of 
the brane tension (the vacuum energy associated with the brane)
and the matter localized on it. The RS model is
the only example in which this has been realized in such
a way that conventional cosmology is reproduced in the
low density limit. However, 
we view this model only as a toy one. Many of its ingredients, such
as the negative energy density in the bulk and the structure of the
resulting AdS space, do not seem crucial for the emergence of accelerating 
solutions. This view is supported by
the discussion of eq. (\ref{sample}) in
the introduction.
In order to test this conclusion, it would be interesting to
implement this mechanism in a model with a bulk geometry that is flat, 
at least asymptotically. 

An important property of the realization of the mechanism 
in the context of the RS model is the absence of acceleration
during radiation domination, even for energy influx. 
For brane matter with
$w=1/3$ the solution (\ref{13a}), (\ref{13b})
describes a radiation dominated brane, whose expansion is decelerating. 
This means that the 
cosmic acceleration is a phenomenon only of the
era of matter domination. 

It must be pointed out that the solution (\ref{13a}), (\ref{13b}) 
depends crucially on the presence of the component $\chi$ of eq. (\ref{chi}).
In the case of the pure AdS bulk space of the 
RS model this component describes,
according to the AdS/CFT correspondence, 
the boundary conformal field theory \cite{gubser}. 
It has been characterized  as ``mirage'' or ``Weyl radiation''.
In our scenario we have assumed a more general matter content for 
the bulk theory, for which this interpretation is not established.  
However, the term $\sim 4H\chi$ in the 
evolution equation of $\chi$ is characteristic of radiation. 
It is not clear if a similar ``radiation'' component will be 
present in a different realization, with an asymptotically flat bulk space
for example. 
Therefore, the absence of acceleration during the radiation dominated
era on the brane, even for energy influx,  may be a property of 
RS-type models only.

The energy influx may not be present at 
all times for other reasons. 
For example, it may be significant only for very low
energy densities on the brane, and be replaced
by energy outflow for densities exceeding a critical value.
This again would lead to the conclusion that
the accelerated expansion is only 
a recent phenomenon in cosmic terms.
The dynamical localization mechanism, that determines
the details of energy accretion, is the main 
direction of further research in order to understand this point and
construct phenomenologically viable models.

\end{document}